\documentclass[12pt]{article}%
\usepackage{amsmath}
\usepackage{amsfonts}
\usepackage{amssymb}
\usepackage{graphicx}%
\providecommand{\U}[1]{\protect\rule{.1in}{.1in}}
\begin{document}
\begin{titlepage}
\begin{flushright}
PUPT-\\
hep-th/yymmnnn
\end{flushright}
\vspace{7 mm}
\begin{center}
\huge{From Quarks to Strings }
\end{center}
\vspace{10 mm}
\begin{center}
{\large
A.M.~Polyakov\\
}
\vspace{3mm}
Joseph Henry Laboratories\\
Princeton University\\
Princeton, New Jersey 08544
\end{center}
\vspace{7mm}
\begin{center}
{\large Abstract}
\end{center}
\noindent
In this article, prepared for the book "The birth of string theory", I recall the sequence of
ideas which led to non-critical strings and gauge/strings duality. I also comment on some
promising future directions.
\vspace{7mm}
\begin{flushleft}
November 2008
\end{flushleft}
\end{titlepage}\bigskip\ 

In the sixties I was not much interested in string theory. The main reason for
that was my conviction that the world of elementary particles should allow
field theoretic description and that this description must be closely
analogous to the conformal bootstrap of critical phenomena. At the time such
views were very far from the mainstream. I remember talking to one outstanding
physicist. When I said that the boiling water may have something to do with
the deep inelastic scattering, I received a very strange look. I shall add in
the parenthesis that this was a beginning of the long series of "strange looks
" which I keep receiving to this day.

Another reason for the lack of interest was actually the lack of abilities. I
could not follow a very complicated algebra of the early works on string
theory and didn't have any secret weapon to struggle with it. On the other
hand, the Landau Institute , to which I belonged, was full of the top-notch
experts in condense matter physics. I remember that in the late sixties to
early seventies Tolya Larkin and I discussed ( many times ) whether
Abrikosov's vortices could be viewed as elementary particles. Nothing concrete
came out of this at that time, but it helped me with my later work. With some
imagination we could have related the vortex lines with strings but we missed it.

I was exploring renormalizible field theory and found the jet structure of
particle production and the sum rules for deep inelastic scattering in all
such theories \cite{pol70}. Then, in the spring of '73 the news about the
asymptotic freedom reached the Landau Institute . In a few days after we (
Sasha Migdal and I ) saw the papers on it (and checked the calculations ), we
had no more doubts that the field theory approach to elementary particles was
the right one. My general formulae worked beautifully in this case.

It was immediately clear that the most important thing now is to study the
non-perturbative phenomena. Here I was helped by the fact that I knew ( and
refereed) the dissertation of Vadim Berezinsky on the 2d spin systems on the
lattice. I generalized his treatment to the case of the non-abelian gauge
fields and developed lattice gauge theory. I didn't expect that other people
were occupied with the same subject. As Ken Wilson recently wrote
\cite{wilson5}:" \emph{If I had not completed and published my work in a
timely fashion then it seems likely that Smit, Polyakov or both would have
produced publications that would have launched the subject".} To that I can
only add that while I indeed had the complete lattice gauge theory at that
time , I lacked one very fundamental element - Wilson's criterion for
confinement, the area law. When Wilson's paper appeared I decided not to
publish my draft before finding something new. It took some more months before
I added to my work the abelian theory of quark confinement based on the idea
of instantons \cite{pol75}. The result was quite stunning - in 3d the
instantons ( which were magnetic monopoles) lead to the formation of the
electric string for all couplings, while in 4d the instantons where the closed
loops of the monopoles trajectories and the confinement occurred after the
coupling exceeded the critical one. A little later Gerard t'Hooft and Stanley
Mandelstam arrived at the qualitative picture of dual superconductors , which
is of course equivalent to the one I just described.

Things were gliding smoothly- it seemed that all is needed was to find the
non-abelian instantons and to look at their interaction. Their disordering
effect would provide the theory of quark confinement, just as in the abelian
case. And indeed the non-abelian instantons have been discovered
\cite{bela75}. More over, we found a very nice self-duality equations for them
and uncovered their topological origin. Many beautiful and important things
were revealed in the following years, the instantons and our self-duality
equation have a big impact on physics and mathematics.

All this was great, but my efforts to build the theory of non-abelian
confinement went nowhere. The reason was that the perturbative effects were
strong in the infrared and could potentially obliterate the instantons. We
know today that at best one can build a reasonable phenomenological theory
based on instantons or , if one looks for the exact theory,one has to escape
to the beautiful countryside of supersymmetric gauge theories in which the
perturbative fluctuations are cancelled or controlled. Indeed, in the nineties
Seiberg and Witten \cite{sei94}managed to guess the exact form of all
instanton corrections in the case of $N=2$ supersymmetry and to discuss the
(essentially abelian) quark confinement in this case. Later Nekrasov gave a
direct derivation, by summing over the instantons \cite{nekr}. So in this
special case the instanton picture was fully proved.

These works had a tremendous impact in various fields but they provided little
help in non-supersymmetric theories. By the end of '77 it was clear to me that
I needed a new strategy and I became convinced that the way to go was the
gauge/ string duality. It made its appearance already in the Wilson work on
the lattice gauge theory, in which the strong coupling expansion was described
as a sum over random surfaces. These surfaces were the result of propagation
of one dimensional objects- electric fluxes. The major difficulty was to find
the continuous limit of this picture. But already on the qualitative level I
found the picture very useful. It helped me to predict the deconfining
transition, leading to the quark- gluon plasma \cite{pol78}. This transition
takes place simply because the strings are melting, as can be seen from the
Peierls argument.

This picture of the strings describing the flux lines is often confused with
the t' Hooft picture which suggests that the string world sheet appears
because the lines of Feynman diagrams become dense. In the normal gauge theory
this certainly doesn't happen. These two pictures are quite different.
However, t'Hooft's estimate of string interaction as $1/N^{2}$ for the SU(N)
theory works in both pictures.

My concrete plan was to write the loop equation for the Wilson loop and then
to represent its solution as a sum over random surfaces. Fortunately I grossly
underestimated the depth and the difficulty of this problem. I managed to
convince Sasha Migdal that the loop equations is the way to go. He joined
forces with Yura Makeenko and they produced an important piece of work
\cite{mac79} .On my side, I also played with the various versions of loop
equations and an idea of "integrability in the loop space" \cite{pol79}. I
also thought that the string representation may help to solve the 3d Ising
model by reducing it to the free fermionic strings ( in 2d it is reduced to
free fermions ).

By then the major challenge became the second part of the program - finding ,
or even defining, the sum over random surfaces. In the case of paths it has
been long known ( since Fock and Schwinger) that it is convenient to use a
quadratic action and then to integrate over the proper time. In the case of
surfaces one can use the analogous quadratic action and introduce the
independent metric on the world sheet. Brink, di Vecchia , Howe and Wess and
Zumino \cite{bdvwz76} ingeniously used this trick to derive supersymmetric
action for string theory. Quadratic action was also used in the twenties in
the famous work by J. Douglas on the Plateau problem.

This action is called now the Polyakov action, demonstrating the Arnold
theorem,stating that things are never called after their true inventors. A
second application of this theorem, to which we are coming now , is the
Liouville action. Namely, I found that there is a crucial difference between
the vibrations of classical and quantum strings. Classically the string is
infinitely thin and has only transverse oscillations. But when I quantized it
there was a surprise - an extra, longitudinal mode, which appears due to the
quantum " thickening" of the string. This new field is called the Liouville
mode. It was very surprising to find that the Lagrangian for this field is
proportional to D-26 or (D-10 in the supersymmetric case), the numbers 26 and
10 were the only dimensions in which the standard string theory has been
formulated. I obtained these numbers after long struggle with ghosts and when
I called Sasha Migdal and told him about this result, he was certain that I
was pulling his leg.

My dream at this point was to use this non-critical string to solve both gauge
theories and the 3d Ising model. Even before finishing the paper I made a one
loop estimate of the critical exponent for the Ising model. I was told by the
experts that it is amazingly close to the experimental value ( which I didn't
remember). Sadly, the last check before sending the paper for publication
showed that I made a mistake, and this "result" was removed from the text.

Still, I was delighted to have a wonderful new playground. I hoped to learn
more not only about gauge theories but also to study two dimensional gravity
on the world sheet as a toy model of real gravity.The fact (due to Sherk and
Schwarz) that the real gravity is a part of string theory added some spice to
the project. This project kept me busy for the next 25 years. It started with
the attempt to build a conformal bootstrap for the Liouville theory. We worked
on it with my friends Sasha Belavin and Sasha Zamolodchikov. We developed a
general approach to conformal field theories, something like complex analysis
in the quantum domain. It worked very well in the various problems of
statistical mechanics but the Liouville theory remained unsolved. I was
disappointed and inclined not to publish our results. Fortunately, my
coauthors had a better judgement than I and our paper turned out to be useful
in a number of fields.

Dynamics of 2d gravity is very rich and even now not completely explored. One
of the problems was the field -dependent cut-off which one must use in order
to preserve general covariance on the world sheet. I tried to overcome this
difficulty by using a different gauge. I found, quite unexpectedly, the
emergence of the SL(2,R) current algebra and ,in a subsequent joint paper by
Sasha Zamolodchikov , Dima Knizhnik and myself, this symmetry allowed us to
find the fractal dimensions of minimal models dressed by the gravitational
field. This work had a tragic element. Dima, my fantastically talented
graduate student, died of the sudden heart failure before the work was done. I
didn't even know that he was working on this subject. But after his death
Sasha and I read his notes and received a crucial insight, which allowed us to
finish the work.

A few years before this work Kazakov and David suggested that the discrete
version of 2d gravity can be described by the various matrix models. It was
hard to be certain that these models really have a continuous limit described
by the Liouville theory, there were no proofs of this conjecture. To our
surprise we found that the anomalous dimensions coming from our theory
coincide with the ones computed from the matrix model. That left no doubts
that in the case of the minimal models the Liouville description is equivalent
to the matrix one. This relation received a lot of attention. Later Witten
found a third description of the same system in terms of the topological field theories.

Another aspect of our theory was a relation between gravity and SL(2,R) gauge
fields. In '89 I wrote : \textit{" It is possible that in this strong gravity
region description in terms of the metric tensor breaks down and gauge fields
should become fundamental variables. If so, we encounter one of the most
exciting situations in physics. " \cite{pol89}}

I kept thinking about gauge/ strings dualities. Soon after the Liouville mode
was discovered it became clear to many people including myself that its
natural interpretation is that random surfaces in 4d are described by the
strings flying in 5d with the Liouville field playing the role of the fifth
dimension. The precise meaning of this statement is that the wave function of
the general string state depends on the four center of mass coordinates and
also on the fifth, the Liouville one. In the case of minimal models this extra
dimension is related to the matrix eigenvalues and the resulting space is flat.

In '96 I came to the conclusion that in order to describe gauge theories this
five dimensional space must be warped. The logic was as following. In gauge/
string duality the open strings describe the Wilson loop and the only allowed
vertex operators in the open string sector are the ones corresponding to
gluons ( and extra fields , if present). At the same time, in the closed
string sector we have infinite number of states. So, all massive modes of the
open string must go away. This can happen only if the ends of the open strings
lie either at singularity or infinity and the metric is such that this region
has infinite blue shift with respect to the bulk. In this case the masses of
all but massless open string states go to infinity.

Since this 5d space must contain the flat 4d subspace in which the gauge
theory resides, the natural ansatz for the metric is just the Friedman
universe with a certain warp factor. This factor must be determined from the
conditions of conformal symmetry on the world sheet. Its dependence on the
Liouville mode must be related to the renormalization group flow. As a result
we arrive at a fascinating picture - our 4d world is a projection of a more
fundamental 5d string theory. As was written 25 centuries ago :

\textit{"They see only their own shadows, or the shadows of one another, which
the fire throws on the opposite wall of the cave."}

A small improvement of Plato - the cave has five dimensions , while the wall - four.

At this point I was certain that I have found the right language for the
gauge/ strings duality. I attended various conferences, telling people that it
is possible to describe gauge theories by solving Einstein-like equations (
coming from the conformal symmetry on the world sheet) in five dimensions. The
impact of my talks was close to zero. That was not unusual and didn't bother
me much. What really caused me to delay the publication (\cite{pol97}) for a
couple of years was my inability to derive the asymptotic freedom from my
equations. At this point I should have noticed the paper of Igor Klebanov
\cite{kle97} in which he related D3 branes described by the supersymmetric
Yang Mills theory to the same object described by supergravity. Unfortunately
I wrongly thought that the paper is related to matrix theory and I was
skeptical about this subject. As a result I have missed this paper which would
provide me with a nice special case of my program. This special case was
presented little later in full generality by Juan Maldacena \cite{mald} and
his work opened the flood gates. The main idea was that for the supersymmetric
Yang-Mills theory the geometry in five dimensions is determined by the
conformal symmetry in the target space. This is the geometry of AdS5 space
which has constant negative curvature. After that , Gubser, Klebanov and I and
Ed. Witten realized that the gauge theory should be placed at infinity in this
space and gave a prescription for calculating various physical quantities.

In order to justify my picture I have used intuition coming from the loop
equation, while Klebanov and Maldacena appealed to the D brane picture of the
gauge fields. Both points of view are useful but neither of them lead to the
quantitative derivation of gauge/string duality.

In the case of D branes the logic is as following. We start from the flat
space and place there a large number of the D branes. Their small oscillations
are described by the large N gauge theory. A nice fact about this
representation of the gauge fields is that it allows us to use geometrical
intuition instead of abstract field theory when considering different
configurations of the D branes. But then one has to take a major step and
postulate that the collection of the D branes can be replaced by their mean
gravitational field . This is a little like replacing the famous cat by its
smile. While this is most probably correct, it is not clear how to justify
this result. Also, there could be cases of gauge/string duality in which the
flat space D branes representation does not exist.

In the case of the loop equations , we argue that since the Wilson loop is
zigzag invariant ( the back and forth parts of the Wilson loop cancel) , the
open string must have the finite number of states (or vertex operators),
corresponding to the states of the gauge theory. As we explained above, this
requirement implies warping needed to remove all higher states from the
spectrum. The approach based on the loop equations starts from the first
principles. However, the equation itself is singular and requires
elaborations. Rychkov and I tried to use the operator product expansions on
the contour in order to find a non-singular version of the equation. We only
scratched the surface of highly non-trivial technical problems. The problem of
reproducing gauge perturbation theory from the string theory side remains
unsolve ( and extremely important).

Why should we care about the derivation from the first principles ? After all,
in physics we value not so much the proved theorems but correct and powerful
statements. However, in this case the lack of the derivation really impedes
progress. We do not know how far the gauge /string duality can be extended and
generalized. The enormous accumulation of special cases has been useful but
not sufficient for deeper understanding. This is why I think that establishing
the foundations is one of the most important problems in the field.

Another important problem is integrability. In the seventies I was very
impressed by the discovery by Belavin and Zakharov \cite{bel} that the
self-duality equations are completely integrable.  I realized \cite{pol79}
that in the case of the full quantum Yang- Mills theory one can expect "
integrability in the loop space". That means that the densities of the
conserved quantities depend not on the points but on the contours. Today, due
to the work of many people, we know that the dilatation operator of the super
- Yang- Mills theory is represented by a completely integrable spin chain. It
is superficially different from the integrability which I envisaged 30 years
ago. However they must be related, since the AdS5 string sigma model is
integrable and the boundary of the world sheet, being mapped onto the Wilson
loop, must produce the integrals in the loop space. Establishing this fact is
one of the yet unsolved problems.

Even more important is to find the gauge theory for the de Sitter space. I
conjectured that the large N gauge theories have a fixed point at the complex
gauge coupling corresponding to the radius of convergence of the planar
graphs. Presumably this point is described by a non-unitary CFT corresponding
to the intrinsically unstable de Sitter space. This approach will hopefully
resolve the puzzle of the cosmological constant and cosmic acceleration.

Also, I confess that I still have some hopes that the 3d critical phenomena
can be approached by string- theoretic methods. The methods of CFT and
holography may also be useful in the problem of turbulence.

As for the problem of string unification, it seems to me that non-critical
strings may have some future. However, it may be wise to wait for some more
information about Nature (specifically about supersymmetry) which we expect to
get from the LHC.

To sum up we have a large number of concrete and fascinating problems which
will entertain us for many years to come. No end of physics in sight.

This work was partially supported by the National Science Foundation grant PHY-0756966.

\end{document}